\documentclass[runningheads]{llncs}
\usepackage{graphicx}
\usepackage{amsmath}
\usepackage{multirow}
\usepackage{amsfonts}
\usepackage[dvipsnames]{xcolor}
\usepackage[linesnumbered,ruled]{algorithm2e}
\begin{document}
\title{VolTS: A Volatility-based Trading System to forecast Stock Markets Trend using Statistics and Machine Learning\thanks{corresponding author: Ivan Letteri}}
%
%
\author{Ivan Letteri\inst{1}\orcidID{0000-0002-3843-386X}}
\authorrunning{I. Letteri}
%
\institute{University of L'Aquila\\
Coppito, snc, 67100 L'Aquila, Italy\\
\email{ivan.letteri@univaq.it}\\
\url{http://www.ivanletteri.it}
}
\maketitle              
\begin{abstract}
Volatility-based trading strategies have attracted a lot of attention in financial markets due to their ability to capture opportunities for profit from market dynamics. In this article, we propose a new volatility-based trading strategy that combines statistical analysis with machine learning techniques to forecast stock markets trend.

The method consists of several steps including, data exploration, correlation and autocorrelation analysis, technical indicator use, application of hypothesis tests and statistical models, and use of variable selection algorithms. In particular, we use the k-means++ clustering algorithm to group the mean volatility of the nine largest stocks in the NYSE and NasdaqGS markets. The resulting clusters are the basis for identifying relationships between stocks based on their volatility behaviour. Next, we use the Granger Causality Test on the clustered dataset with mid-volatility to determine the predictive power of a stock over another stock. By identifying stocks with strong predictive relationships, we establish a trading strategy in which the stock acting as a reliable predictor becomes a trend indicator to determine the buy, sell, and hold of target stock trades. 

Through extensive backtesting and performance evaluation, we find the reliability and robustness of our volatility-based trading strategy. The results suggest that our approach effectively captures profitable trading opportunities by leveraging the predictive power of volatility clusters, and Granger causality relationships between stocks.

The proposed strategy offers valuable insights and practical implications to investors and market participants who seek to improve their trading decisions and capitalize on market trends. It provides valuable insights and practical implications for market participants looking to.

\end{abstract}
\section{Introduction}
\label{sect:intro}

Volatility-based trading is an increasingly important area in the financial industry as it offers opportunities to capitalize on market dynamics. Artificial intelligence (AI) is bringing volatility-based trading to the market, playing an important role to create robust tools and techniques for analyzing and leveraging market volatility. In particular, using AI to estimate mean volatility provides valuable insight into the uncertainty and risk associated with certain securities or the overall market \cite{letteri2022dnnforwardtesting}.

Moreover, intraday volatility forecasts are important for risk management \cite{jjfinecnbr005}. Additionally, volatility forecasts help traders assess the likelihood of price fluctuations and help understand automated trading strategies \cite{bates2019}, increasingly managed by AI. In particular, the application of statistical methods and machine learning techniques has in recent years provided a new approach to the development of innovative and profitable trading strategies \cite{MLPletteriStockTrading}.

To the best of our knowledge, unlike the forecasting of daily volatility, the research literature has paid little attention to volatility \cite{ecofch115} but does not adequately capture specific characteristics of intraday returns.

The main objective of this study is to develop an AI trading strategy using the clustering of average volatility data, calculated using the k-means++ algorithm \cite{kmeans++}, from a set of nine major stock markets. The initial aim is to identify different volatility regimes that exist in the market and group assets based on these regimes. We then apply the Granger Causality Test (GCT) \cite{Kirchgassner2007} to identify stocks that act as significant predictors of other stocks in our analysis universe to set targets for buying, selling, or holding trades.

Through our AITA framework \cite{aita2023}, we conducted a thorough empirical analysis to assess the return and performance of the proposed strategy using historical data. Various performance evaluation metrics were considered to assess the effectiveness and robustness of the strategy in generating profits.

The literature on technical trading strategies is quite robust, with most studies tending to focus on moving averages of prices as the relevant indicators for determining when to enter/exit investments \cite{LetteriFEMIB23}. We examine the Historical Volatility (HV) estimators as a potential dataset from which to extrapolate the medium volatility to select stocks to apply the GCT for asset cointegrations approach \cite{Engle1987}.

The following sections of this paper will focus on the details of our proposed method. Section \ref{sect:prel} reports preliminary concepts about the AITA framework and the specific modules involved in this research (\textit{VolTS} and \textit{AitaBT}). Section \ref{sect:method} describes the methodology integrated into the AITA module called \textbf{VolTS} (\textit{Volatility Trading System}) to group the volatility average of the securities and identify the predictive relations between them, followed by the experiment of the trading strategy based on these results, with in-depth empirical analysis to assess the performance and robustness of the strategy. In sections \ref{sect:results} the practical results reported with the backtesting using the AitaBT module, and the discussion. Finally, section \ref{sect:conclusion} concludes this study by summarizing with a clear perspective on the effectiveness and applicability of the proposed method.

\section{Background about AITA framework}
\label{sect:prel}
\subsection{AITA Price Action module}
\label{sec:techAnalysis}
The price action (PA) influences HV, and in turn, HV can provide insights into future PA. When the PA exhibits strong price movements, such as wide trading ranges, breakouts, or rapid directional changes, it tends to increase. 
VolTS is the AITA module which follows these rules: low HV indicates a period of consolidation or low price volatility, suggesting a potential upcoming volatility spike or a change in the price action. On the other side, high HV indicates a higher likelihood of sharp movements or trend changes in the market.

\begin{figure}[!ht]
	\centerline{\includegraphics[width=30em]{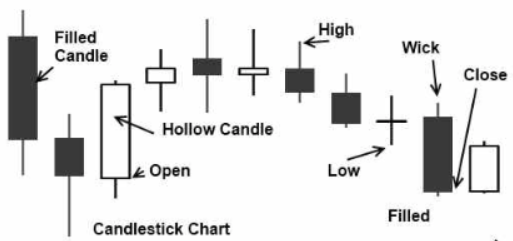}}
	\caption{Example of candlestick chart.}
	\label{fig:candlestick}
\end{figure}

Into VolTS, the PA is encoded as OHLC, i.e., the open, high, low, and close prices of the assets, like represented in the candlesticks charts (see Fig. \ref{fig:candlestick}). For each timeframe $t$, the OHLC of an asset is represented as a 4-dimensional vector $X_t = (x^{(o)}_t,x^{(h)}_t,x^{(l)}_t,x^{(c)}_t)^T$, where $x_t^{(l)} > 0$, $x_t^{(l)} < x_t^{(h)}$ and  $x_t^{(o)}, x_t^{(c)} \in [x_t^{(l)},x_t^{(h)}]$.

\subsection{VolTS Historical Volatility module}
\label{sect:hvs}
The construction of the dataset is designed to use the following HV estimators:


\begin{itemize}
    \item[-] The \textit{Parkinson} (PK) estimator incorporates the stock's daily \textit{high} and \textit{low} prices as follow: $$PK = \sqrt{\frac{1}{4Nln2}\sum_{i=1}^N(ln\frac{x_t^{(h)}}{x_t^{(l)}})^2}.$$ It is derived from the assumption that the true volatility of the asset is proportional to the logarithm ($ln$) of the ratio of the high $x_t^{(h)}$ and low $x_t^{(l)}$ prices of $N$ observations.
    \item[-] The \textit{Garman-Klass} (GK) estimator is calculated as follows: $$\sqrt{\frac{1}{N}(\sum_{i=1}^N \frac{1}{2}(ln\frac{x_t^{(h)}}{x_t^{(l)}})^2-\sum_{i=1}^N(2ln(2)-1)(ln\frac{x_t^{(c)}}{x_t^{(o)}})^2)}.$$ This estimator assumes that price movements are log-normally distributed, which may not always be the case in practice.
    \item[-] The \textit{Rogers-Satchell} (RS) estimator uses the range of prices within a given time interval as a proxy for the volatility of the asset as follows: $$RS = \sqrt{\frac{1}{N}\sum_{t=1}^{N}(ln(\frac{x_t^{(h)}}{x_t^{(c)}})ln(\frac{x_t^{(h)}}{x_t^{(o)}})+ln(\frac{x_t^{(l)}}{x_t^{(c)}})ln(\frac{x_t^{(l)}}{x_t^{(o)}})}.$$ RS assumes that the range of prices within the interval is a good proxy for the volatility of the asset, additionally, the estimator may be sensitive to outliers and extreme price movements.
    \item[-] The \textit{Yang-Zhang} (YZ) estimator \cite{Yang2000Zhang} incorporates OHLC prices as follows: 
    
    $$YZ = \sqrt{\sigma_{OvernightVol}^2 +k\sigma_{Open-to-CloseVol}^2+(1-k)\sigma^2_{RS}},$$ 
    
    where $k = 0.34/1.34+\frac{N+1}{N-1}$, $\sigma_{Open-to-CloseVol}^2 = \frac{1}{N-1}\sum_{i=1}^N(ln\frac{x_t^{(c)}}{x_t^{(o)}} - ln\frac{x_t^{(c)}}{x_t^{(o)}})^2$, and $\sigma_{OvernightVol}^2 = \frac{1}{N-1}\sum_{i=1}^N(ln\frac{x_t^{(o)}}{x_{t-1}^{(c)}} - ln\frac{x_t^{(o)}}{x_{t-1}^{(c)}})^2$.
    
    Empirical investigations have revealed that the YZ estimator manifests commendable performance across a wide range of scenarios, including those featuring jumps and non-normality in the data. Nevertheless, this estimator is not infallible, and its efficacy may be restricted in certain contexts. 
\end{itemize} 
In this study, our focus is on the mid-volatility to either close open positions or avoid entering a position when the expected volatility coefficient is high, thereby limiting the risk of losses. On the other hand, if the expected volatility is too low, it does not present any opportunities for gains.

As a result, employing multiple estimators and comparing their outcomes is frequently recommended to obtain a more holistic comprehension of the latent volatility aspects.

\subsection{AITA Strategies module}
\label{subsect:tsu}
Three distinct trading strategy classes are implemented in AITA framework: 
\begin{itemize}
    \item[-] \textit{Buy and Hold} (B\&H) strategy is used as a benchmark to compare the performance of the two strategies below. It involves buying one single share on the first date of the period studied on the market close and selling the share at the market close on the last date as follows: $V_t = Q \cdot P_t$, where $V_t$ is the value of the investment at time $t$. $Q$ is the quantity of the asset purchased at time $t=0$, and $P_t$ is the price of the asset at time $t$ with $P_0$ the initial price.
    \item[-] \textit{Trend Following} (TF) strategy is one way to engage in trend trading, where a trader initiates an order in the direction of the breakout after the price surpasses the resistance line as follows: let $P_t$ the price at time $t$, and let $MA$ denote the Moving Average of the asset price over a certain period. If $P_t \geq MA_t$ indicates an upward trend to take a long position otherwise it is a downward trend to take a short position.   
   \item[-] \textit{Mean Reversion} (MR) strategy suggests that a security's maximum and minimum prices are temporary, and the security will eventually move towards its mean as follows: let $P_t$ the price of the asset at time $t$, and let $\mu$ and $\sigma$ represent the mean and standard deviation of the asset price, respectively. The entry/exit conditions for a long/short position are given by: $P_t < \mu - k \cdot \sigma$ and $P_t > \mu - k \cdot \sigma$, respectively where $k$ is a constant representing the number of standard deviations from the mean at which the entry condition is triggered.
\end{itemize}

For the sake of brevity, in this study, the focus is on the trend-follow strategy and we compare it with the B\&H considered as a benchmark. It is important to note that both trend-following and mean reversion strategies, which are theoretically opposing concepts, can be applied to the same stock without conflicting with each other. Nonetheless, we find it beneficial to apply the mean reversion strategy when dealing with mid-volatility assets.

\subsection{AitaBT the backtesting module}
AitaBT module considers both profit and risk metrics as crucial factors in trading, in order to evaluate the potential profitability of investments and manage risk exposure.
\begin{itemize}
    \item(i) The \textit{Maximum drawdown (MDD)} measures the largest decline from the peak in the whole trading period, to show the worst case, as follows: $MDD=max_{\tau \in (0,t)}[max_{t \in (0,\tau)}\frac{n_t-n_{\tau}}{n_t}]$. 
    \item(ii) The \textit{Sharpe ratio (SR)} is a risk-adjusted profit measure, which refers to the return per unit of deviation as follows: $SR = \frac{\mathbb{E}[r]}{[r]}$. 
    \item(iii) The \textit{Sortino ratio (SoR)} is a variant of the risk-adjusted profit measure, which applies DD as risk measure: $SoR = \frac{\mathbb{E}[r]}{DD}$.
    \item(iv) The \textit{Calmar ratio (CR)} is another variant of the risk-adjusted profit measure, which applies MDD as risk measure: $CR = \frac{\mathbb{E}[r]}{MDD}$.
\end{itemize}
 
To check the goodness of trades, we mainly focused on the \textit{Total Returns} $R_{k}(t)$ for each stock $(k = 1, ...,p)$ in the time interval $(t= 1, ...,n)$, where $TR = R_{k}(t) = \frac{Z_k(t+\Delta t) - Z_{k}(t)}{Z_{k}(t)}$, and furthermore analysing the standardized returns $r_k = (R_k - \mu_k) / \sigma_k,$ with $(k = 1, ...,p)$, where $\sigma_k$ is the standard deviation of $R_k$, e $\mu_k$ denote the average overtime for the studied period.

\section{Methodology}
\label{sect:method}

\subsection{Assets Selection}
For the experiment, we set the AITA framework to work with the main nine stocks (listed in Tab. \ref{tab:stocks}) from NYSE and NasdaqGS at the moment. It downloads automatically the OHLC prices, via an internal Python library connected to API, using the MetaTrader5 (MT5)\footnote{https://www.metatrader5.com/} directly associated with the broker TickMill \footnote{https://tickmill.eu}.

\begin{table}[!th]
\centering
\caption{List of the main 9 stocks selected for the experimentation.}
\label{tab:stocks}
\begin{tabular}{|l|l|l|}
\hline
\multicolumn{1}{|c|}{\textbf{Ticker}} & \multicolumn{1}{c|}{\textbf{Company}} & \multicolumn{1}{c|}{\textbf{Market}} \\ \hline
MSFT                                 & \multicolumn{1}{c|}{Microsoft Corporation}        & \multicolumn{1}{c|}{NasdaqGS}             \\ \hline
GOOGL                                   & \multicolumn{1}{c|}{Alphabet Inc.}       & \multicolumn{1}{c|}{NasdaqGS}             \\ \hline
MU                                  & \multicolumn{1}{c|}{Micron Technology, Inc.}        & \multicolumn{1}{c|}{NasdaqGS}             \\ \hline
NVDA                                  & \multicolumn{1}{c|}{NVIDIA Corporation}        & \multicolumn{1}{c|}{NYSE}             \\ \hline
AMZN                                   & \multicolumn{1}{c|}{Amazon.com, Inc.}                       &   \multicolumn{1}{c|}{NYSE}                                    \\ \hline
META                                    &  \multicolumn{1}{c|}{Meta Platforms, Inc.}                                     & \multicolumn{1}{c|}{NYSE}                                      \\ \hline
QCOM                                   &  \multicolumn{1}{c|}{QUALCOMM Incorporated}                             & \multicolumn{1}{c|}{NasdaqGM}                                     \\ \hline
IBM                                   &  \multicolumn{1}{c|}{Int. Business Machines Corp.}                                     & \multicolumn{1}{c|}{NYSE}                                     \\ \hline
INTC                                   &   \multicolumn{1}{c|}{Intel Corporation}                                    & \multicolumn{1}{c|}{NYSE}             \\ \hline
\end{tabular}
\end{table}

\subsection{AITA Anomaly Detection} 
AITA framework to default, examines the price time series of the assets to determine the time window without considerable anomalies. The criterion implemented is based on the anomaly score calculated by a K-Nearest Neighbors (KNN) model \cite{Wahid2020Abdul}. One of the key advantages of KNN is its ability to handle non-linear and complex relationships between data points \cite{letteriG1no}\cite{letteriG1noArXiv}. The KNN model is fit to the time series data and the anomaly score is calculated based on the distance between the points and its $k$ nearest neighbours.

The threshold ($th$) for detecting anomalies is then determined based on the mean ($\mu$) and standard deviation ($\sigma$) of the anomaly scores. The criterion can be expressed as follows: let $x_t$ be the value of the time series at time $t$, and $k$ be the number of nearest neighbours to use in the KNN model. With $d_{euc}(x_t, x_i)$, we define the Euclidean distance between $x_t$ and $x_i$, where $x_i$ is the $i^{th}$ nearest neighbor of $x_t$. The anomaly score ($score_t$) for $x_t$ is defined as follow: 
$$ score_t = \frac{1}{k} \sum(d_{euc}(x_t, x_i), \forall i \in NearestNeighbors(x_t, k) .$$


The threshold $th$ for detecting anomalies as follows: $th = \mu + 3 * \sigma$. Data points with anomaly scores greater than the threshold are considered to be anomalies.
\begin{figure*}[!ht]
	\centerline{\includegraphics[width=40em]{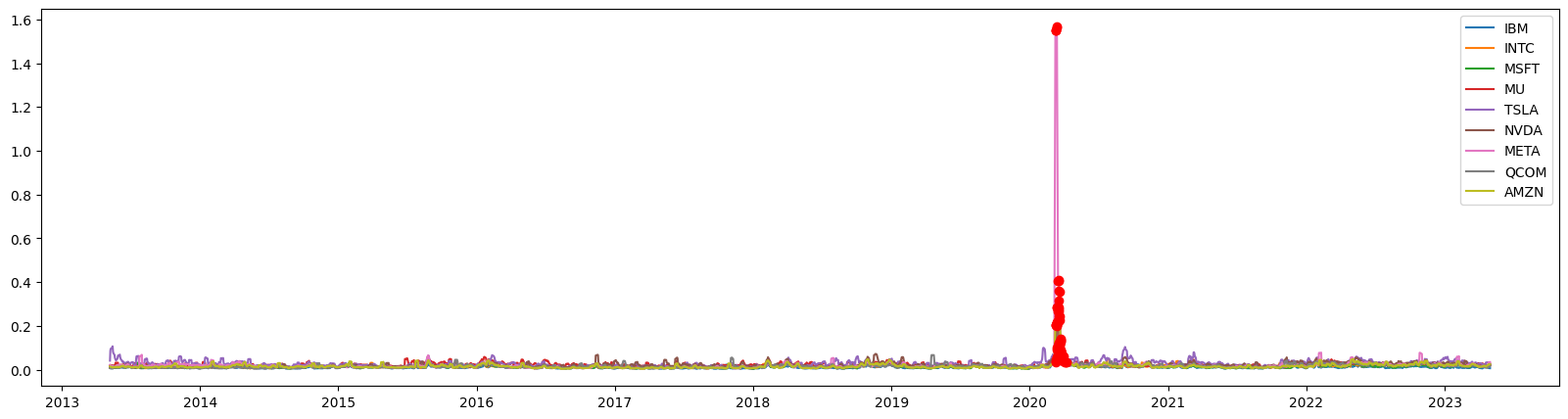}}
	\caption{Red dots highlight the anomalies detected in the interval analyzed from 2020/05/01 to 2023/05/01.}
	\label{fig:anomaly_volatility}
\end{figure*}

Fig. \ref{fig:anomaly_volatility} shows only one critical anomaly during March 2020 (the global pandemic), so we decide to use only the time window in the period after instead to remove it, starting from 1st May 2020 to 1st May 2023.

\subsection{Dataset of Historical Volatility}
The History Volatility Clustering process of our approach determines the stocks with intermediate volatility. First calculate the average of historical volatility time series among the aforementioned estimators (see sect. \ref{sect:hvs}). Next, the resulting volatility series are clusterized using the KMeans++ algorithm with the dynamic time warping (DTW) metric \cite{Niennattrakul2007Vit}. DTW is used to compare couples of time series that may have different lengths and speeds of variation, which makes it well-suited for this type of clustering. In particular, we split into three clusters ($K = 3$) \textit{high}, \textit{middle}, and \textit{low} volatility. The centroids are selected using the maximum DTW distance with respect to the previous centroid. 

\begin{figure*}[!ht]
	\centerline{\includegraphics[width=38em]{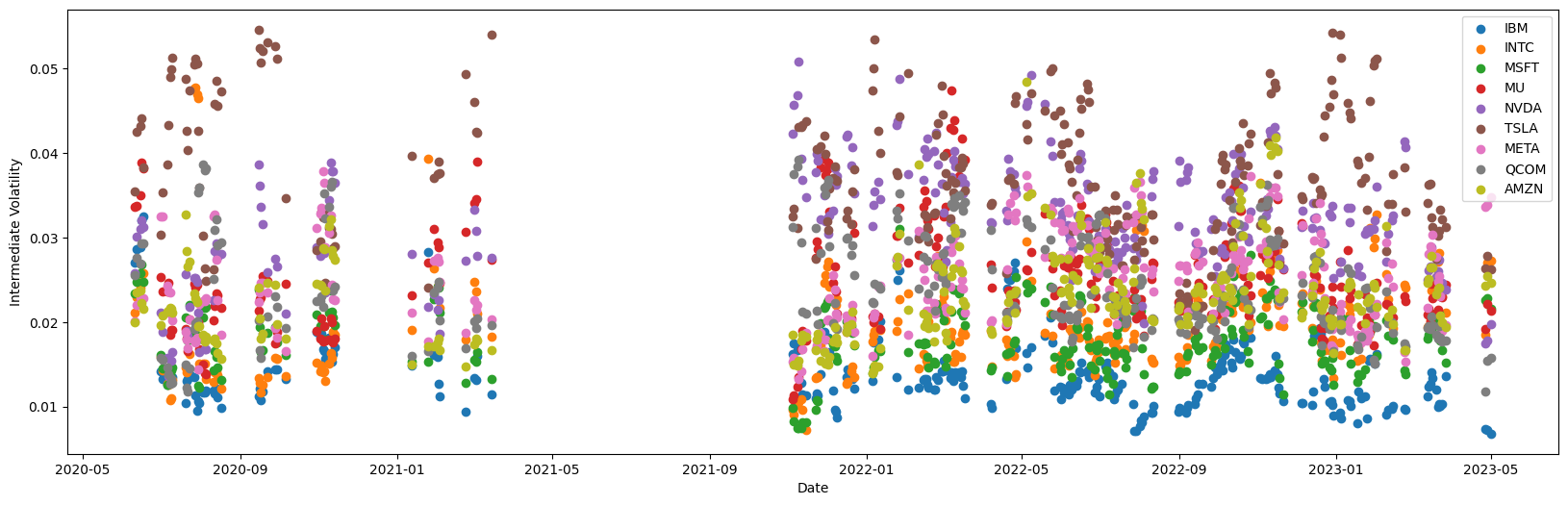}}
	\caption{Kmeans++ clusters with $k=3$ of the Historical Volatility estimators dataset, from 1st May 2020 to 1st May 2023. }
	\label{fig:kmeanClusters}
\end{figure*}

Fig. \ref{fig:kmeanClusters} shows the results displayed through a plot of the time series belonging to the middle cluster where we are focused on our strategy. It is worth noting that, the main region is in the time window from 1st November 2022 to 1st May 2023. So, we use this interval as the dataset, and then from the intermediate cluster, the candidate assets selected are TSLA with the highest, AMZN and META in the middle, with QCOM and IBM with the lowest values, respectively.






In the context of volatility-based trading, VolTS module performs the GCT to examine the relationship between lagged volatility of one asset and the future volatility of another asset by applying the following steps:
\begin{itemize}
    \item \textbf{Step 1.} \textit{Significant Granger causality:} Let $X$ and $Y$ be the pair stocks time series volatility to check, where $X$ represents the potential causal variable and $Y$ represents the potential effect variable. The null hypothesis (H0) states that $X$ does not Granger cause $Y$, while the alternative hypothesis (H1) states that $X$ does Granger cause $Y$. The F-test is:
    $$F=\frac{[(RSS_{Y(t)}-RSS_{YX_{(t)}})/p]}{[RSS_{YX_{(t)}})/(n-p-k)]},$$ 
    where $RSS$ is the \textit{Residual Sum of Squares} for the two AR models:
    $$Y(t)=c_Y+\beta_{Y_{1}}*Y(t-1)+\beta_{Y_{2}}*Y(t-2)+\cdots+\beta_{Y_{p}}*Y(t-p)+\epsilon_{Y(t)},$$ 
    $$X:Y(t)=c_{YX}+\beta_{YX_{1}}*X(t-1)+\beta_{YX_{2}}*X(t-2)+\cdots+\beta_{YX_{p}}*X(t-p)+\epsilon_{YX(t)},$$
    with $p$ the lag order, $n$ the number of observations, and $k$ the number of parameters in the models.\\
    \item \textbf{Step 2.} \textit{F-statistic comparison} with the critical value from the F-distribution where the significance level has $\alpha=0.05$. If the F-statistic is greater than the critical value, reject the null hypothesis (H0) and conclude that $X$ Granger causes $Y$ with statistical significance. If the F-statistic is not greater than the critical value, fail to reject the null hypothesis (H0) and conclude that there is no significant Granger causality between $X$ and $Y$.\\
    \item \textbf{Step 3.} \textit{Direction of causality:} If the volatility of Stock $X$ Granger causes the volatility of Stock $Y$, it suggests that changes in Stock X's volatility can be used to predict changes in Stock Y's volatility.\\
\end{itemize}

\begin{figure}[]
	\centerline{\includegraphics[width=30em]{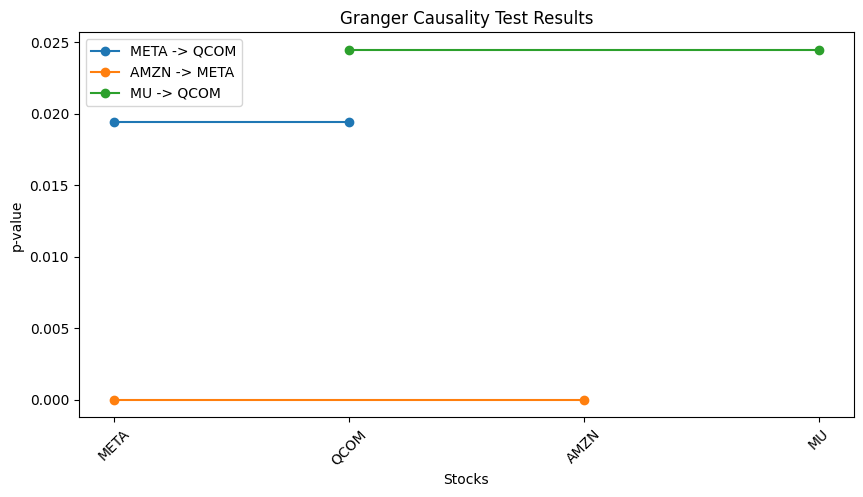}}
	\caption{Co-integration via GCT.}
	\label{fig:cointegrationGCT}
\end{figure}

\begin{figure}[!ht]
	\centerline{\includegraphics[width=20em]{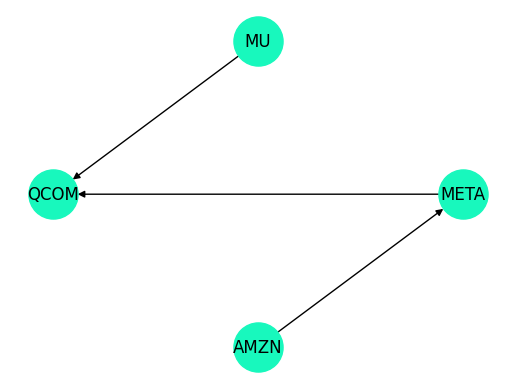}}
	\caption{The best Acyclic Graph of the co-integration.}
	\label{fig:graphGCT}
\end{figure}

The VolTS algorithm (see pseudo-code in Appendix \ref{algo:gct}) iterates the daily lags in a range from 2 to 30 days to determine the best result. In this experiment, the best result is achieved with lags=5, where '\textit{best}' is considered when there is direction coherency among the stocks with the maximum cardinality of the set of stocks. In other words, the GCT direction does not generate the acyclic graph in the connection among the highest number of nodes, as shown in fig. \ref{fig:graphGCT}.

\begin{figure*}[ht]
	\centerline{\includegraphics[width=40em]{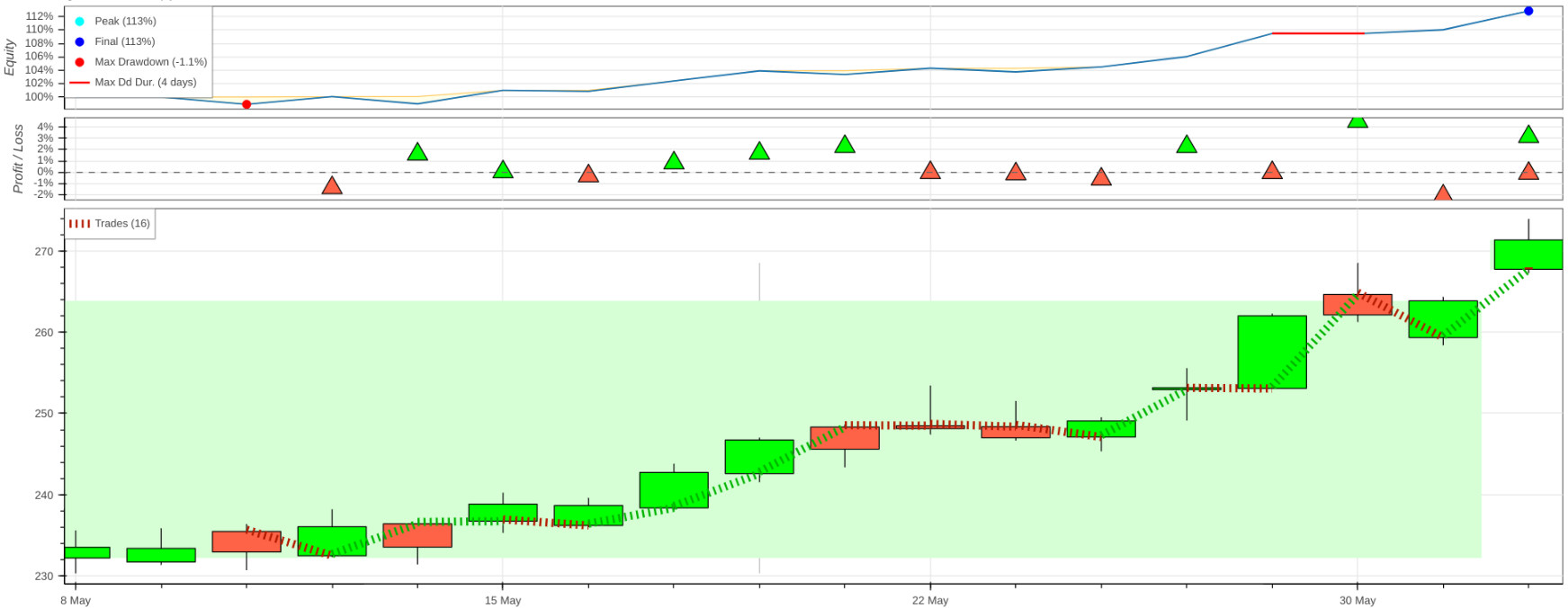}}
	\caption{Co-integration AMZN to META without spurious correlation.}
	\label{fig:meta_plot}
\end{figure*}

In fig. \ref{fig:meta_plot}, we can see how the GCT suggests buying QCOM when META has a positive trend and vice versa, the same thing with MU. Furthermore, when AMZN price increases, it is time to buy META and so on. 

\section{Results and Discussion}
\label{sect:results}


The results of the experiment indicate that the volatility-based trading strategy has performed well during the tested period from 8th April 2023 to 1st June 2023. The strategy resulted in a total gain of 231.77\$ in 40 days of market opening, starting with an initial budget of 1000\$ per stock. The exposure time of the positions being open was quite high at 88.89\% for all the stocks, indicating active trading and frequent changes in the portfolio.

Tab. \ref{tab:BT-stats} contains further details about the performance metrics of the strategy and shows how the total amount in the portfolio is increased to 3231.77\$ (7.725\%), which is a positive sign of profitable trading, also considering the fixed commission of 9\$ per trade. Notice that, the managing of the budget is set in compounded mode, so the full amount is reused for each trade.

\begin{table}[]
\centering
\caption{Results of the backtesting of the experiment.}
\label{tab:BT-stats}
\begin{tabular}{|c|c|c|c|ccc|c|}
\hline
\multirow{2}{*}{\textbf{Stock}}  & \multirow{2}{*}{\textbf{\begin{tabular}[c]{@{}c@{}}Num. of \\ trades\end{tabular}}} & \multirow{2}{*}{\textbf{\begin{tabular}[c]{@{}c@{}}Win \\ rate(\%)\end{tabular}}} & \multirow{2}{*}{\textbf{\begin{tabular}[c]{@{}c@{}}Total \\ return(\$)\end{tabular}}} & \multicolumn{3}{c|}{\textbf{Ratios}}                                                           & \multirow{2}{*}{\textbf{MDD(\%)}} \\ \cline{5-7}
                                 &                                                                                     &                                                                                   &                                                                                       & \multicolumn{1}{c|}{\textbf{Sharpe}} & \multicolumn{1}{c|}{\textbf{Sortino}} & \textbf{Calmar} &                                   \\ \hline
\textit{AMZN -\textgreater META} & 16                                                                                  & 37.5                                                                              & 1045.01                                                                               & \multicolumn{1}{c|}{1.1784}          & \multicolumn{1}{c|}{4.6421}           & 14.264          & 1.77                              \\ \hline
\textit{META -\textgreater QCOM} & 16                                                                                  & 43.75                                                                             & 1110.11                                                                               & \multicolumn{1}{c|}{3.8511}          & \multicolumn{1}{c|}{44.4613}          & 248.342         & -1.33                             \\ \hline
\textit{MU -\textgreater QCOM}   & 16                                                                                  & 56.25                                                                             & 1076.65                                                                               & \multicolumn{1}{c|}{1.2130}          & \multicolumn{1}{c|}{6.3624}           & 23.687          & -7.65                             \\ \hline
\end{tabular}
\end{table}

\begin{figure*}[ht]
	\centerline{\includegraphics[width=40em]{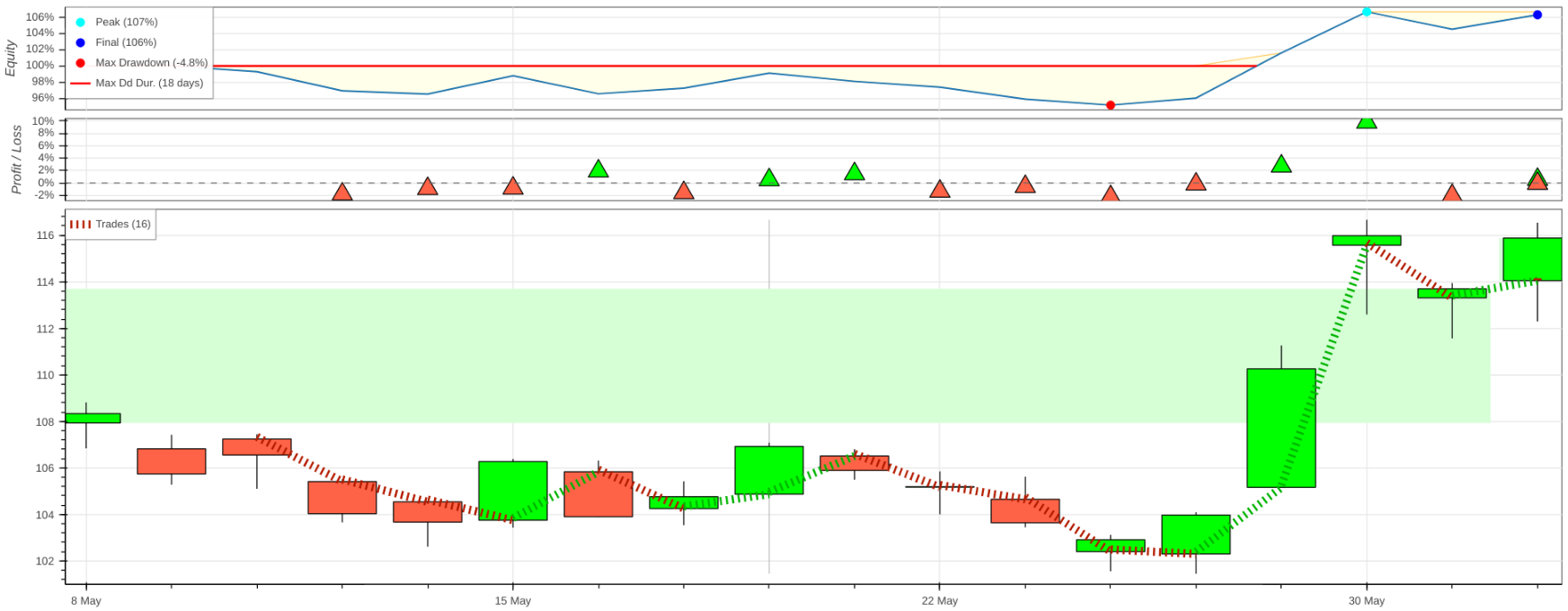}}
	\caption{Co-integration META to QCOM without spurious correlation.}
	\label{fig:qcom_plot}
\end{figure*}

The analysis of individual stocks' performance is presented in Fig. \ref{fig:qcom_plot} about META co-integration. The trades of META bought following the AMZN trend resulted in a Profit and Loss (PnL) of 1.281\%, with a return of 9.721\%. This return outperforms the B\&H strategy, which would have yielded a return of 6.684\%.

\begin{figure*}[ht]
	\centerline{\includegraphics[width=40em]{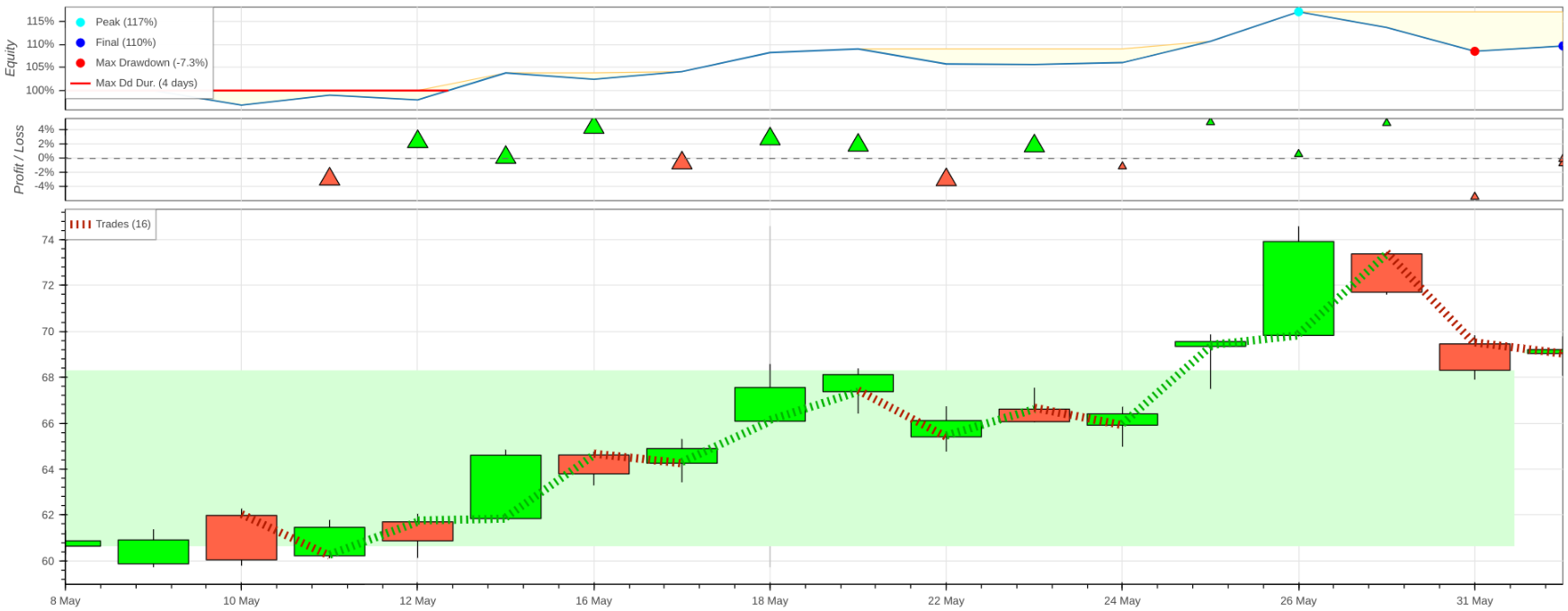}}
	\caption{Co-integration MU to QCOM without spurious correlation.}
	\label{fig:mu_plot}
\end{figure*}

Fig. \ref{fig:mu_plot} shows the trades of QCOM bought following the META trend showed a PnL of 2.774\%, with a return of 12.866\% compared to the B\&H return of 9.235\%. Lastly, the trades of QCOM bought following the MU trend resulted in a PnL of 1.562\%, with a return of 6.302\% as opposed to the B\&H return of 3.969\%.

These results suggest that the volatility-based trading strategy, guided by the Granger Causality Test combined with the moving average trend following strategy, has been successful in generating profits for the selected stocks during the given period. The strategy's active approach and adaptive decision-making based on the trends of other stocks have proven to be more effective than the simple B\&H strategy for the stocks in question. However, it is essential to consider that past performance is not indicative of future results, and further testing and validation would be necessary to assess the strategy's robustness and reliability under different market conditions.

\section{Conclusion and Future Direction}
\label{sect:conclusion} 
   
In this work, we propose an effective method to handle volatility in trading strategy and combine causality by the GCT implemented in the AITA framework with the module VolTS.

In such a system, the trades are guided by co-integration applied to a set of pre-selected stocks, as usual in algorithmic trading. However, the novelty of the presented approach is in the selection of the mid-volatility assets and the use of GCT on the historical market data, and choosing the most profitable pairing between stocks. On the other hand, in our approach, the predictive property is chosen by K-means++ combined with a statistical method. 

Given the promising potentials of this approach, we will further test its reliability on other stock markets using different data, such as cryptocurrencies or defi-tokens, also varying the timeframes for day trading and scalping activities.

We offer several potential areas of future research: (i) exploring more powerful techniques to filter and integrate text information using datasets optimization approaches \cite{letteri2020DOS}\cite{LetteriCP2021intellisys}, and (ii) incorporating domain expert knowledge to enhance model comprehension of price and volume information.

Furthermore, we will expose the API of the AITA framework as a Service considering security methodology to prevent botnet attacks using Deep Learning models \cite{LetteriPG19SecIOT}\cite{LetteriPC19HTTP}. For resilience, we plan to create a Multi-Agent System, also from a perspective of ML for transparent Ethical Agents for customer service \cite{DyoubCLL20} with an evaluation of dialogues \cite{Letteri2109ethMon} using the supervision of an ethical teacher \cite{DyoubCL22ethTeach}. 

%
%
%
\bibliographystyle{splncs04}
\bibliography{bibliography}
%





\newpage

\section*{Appendix}
\begin{algorithm}[H]
\SetAlgoLined
\SetKwInOut{Input}{Input}
\SetKwInOut{Output}{Output}
\Input{$x^{(c)}_t$\_TS: DataFrame}
\Output{selected\_stocks: set}
\BlankLine
lag\_days $\gets 2$\;
\tcp{Step 1: Iteration from 2 to 30}
\While{lag\_days$ \neq 31$}{
\BlankLine
\tcp{Step 2. F-statistic comparison}
granger\_results $\gets$ [];\\
threshold $\gets$ 0.025;

\For{stock1 \textbf{in} $x^{(c)}_t$\_TS.columns}{
\For{stock2 \textbf{in} $x^{(c)}_t$\_TS.columns}{
\If{stock1 \textbf{is equal to} stock2}{
\textbf{continue};
}
data $\gets$ concat([$x^{(c)}_t$\_TS[stock1], $x^{(c)}_t$\_TS[stock2]]);\\
data.columns $\gets$ [stock1, stock2];\\
result $\gets$ GCT(data, maxlag=lag\_days);\\
p\_value $\gets$ result[lag\_days][0]['ssr\_ftest'][1];\\
\If{p\_value \textbf{is less than} threshold}{
granger\_results.append((stock1, stock2, p\_value));
            }
        }
    }
    lag\_days $\gets$ ++;
}
\BlankLine
\tcp{Step 3. Direction of causality}
selected\_stocks $\gets$ set();\\
\For{(stock1, stock2, \_) \textbf{in} granger\_results}{
selected\_stocks.add(stock1);\\
selected\_stocks.add(stock2);
}
\BlankLine
\Return{selected\_stocks};
\caption{Algorithmic strategy via GCT.}\label{algo:gct}
\end{algorithm}

\end{document}